\documentclass[preprint,showpacs,preprintnumbers,amsmath,amssymb, superscriptaddress]{revtex4}

\usepackage{textcomp}
\usepackage{makeidx}
\usepackage{amsmath}
\usepackage{subfigure}
\usepackage{amssymb}
\usepackage{hyperref}
\usepackage{cleveref}
\usepackage{graphicx}
\usepackage[utf8]{inputenc}
\usepackage{float}
\usepackage{color}
\usepackage{epsfig}
\usepackage{eucal}
\usepackage{amsmath}
\usepackage{tabularx}
\usepackage{tabulary}
\usepackage{graphics}
\usepackage{blindtext}
\usepackage{dblfloatfix}
\begin{document}

\title{Compact Anisotropic Models in General Relativity by Gravitational Decoupling}

\author{E. Morales}
\email{emc032@alumnos.ucn.cl}
\affiliation{Departamento de F\'isica, Universidad Cat\'olica del Norte, Av. Angamos $0610$, Antofagasta, Chile.}
\author{Francisco Tello-Ortiz}
\email{francisco.tello@ua.cl}
\affiliation{Departamento de F\'isica, Facultad de ciencias básicas, Universidad de Antofagasta, Casilla 170, Antofagasta, Chile.}


\begin{abstract}
 Durgapal's fifth isotropic solution describing spherically symmetric and static matter distribution
 is extended to an anisotropic scenario. To do so we employ the gravitational decoupling through the minimal geometric deformation scheme. This approach allows to split Einstein's field equations in two simply set of equations, one corresponding to the isotropic sector and other to the anisotropic sector described by an extra gravitational source. The isotropic sector is solved by the Dugarpal's model and the anisotropic sector is solved once a suitable election on the minimal geometric deformation is imposes. The obtained model is representing some strange stars candidates and fulfill all the requirements in order to be a well behaved physical solution to the Einstein's field equations.
\end{abstract}

\pacs{}

\keywords{}
\maketitle

\section{Introduction}
In 1915 Albert Einstein \cite{einstein,einstein1,einstein2,einstein3} stunned the scientific community by presenting one of the greatest achievements of theoretical physics and at the same time one of the most beautiful theories known to date, which also has a great experimental support \cite{Will}. We are talking about General Relativity (GR from now on). Shortly after its publication, Schwarzschild \cite{schwar} presented the first solution to the Einstein field equations. Solution that describes the neighborhood of a compact object that is spherically symmetric and static and free of material content, i.e. vanishing pressure and density. Later decades witnessed the great development that this wonderful theory experienced. Proof of this is reflected in the work of Professor Tolman \cite{tolman}, who found several analytical solutions which describe stellar interiors subject to a spherically symmetric and static geometry whose matter distribution corresponds to a perfect fluid (equal radial and tangential pressures $p_{r}=p_{t}$). However, subsequent investigations showed that configurations with spherical symmetry do not necessarily meet the condition $p_{r}=p_{t}$. Of course, the introduction of anisotropies (unequal radial and tangential pressures, $p_{r}\neq p_ {t}$) allowed a better understanding about highly dense objects. The pioneering works by Ruderman \cite{ruderman} and Bowers and Liang \cite{bowers} about anisotropic matter distribution opened the doors to one of the most researched branches nowadays. At present it remains a great challenge to find solutions to the Einstein field equations that also meet the requirements of physical admissibility \cite{delgaty}. Despite the great difficulty that it presents, an extensive variety of works available in the literature \cite{mehra,Ivanov, harko1, schunk, usov, deb, monadi, deb1, shee, rahaman4, rahaman, rahaman1, rahaman3, rahaman2, varela, negreiros, mello, ray, kileba1, hassan1, maurya9, maurya, maurya1, maurya2, maurya3, maurya4, maurya5, maurya8, maurya6, maurya7, kileba, hassan,bhar1, newton, newton1, newton2, newton3, newton4, newton5, jasim,smaurya,smaurya1,smaurya2} (and reference
contained therein) have successfully addressed the study and understanding of the role played by the anisotropy in stellar interiors. As Mak and Harko have argued \cite{harko1}, anisotropy can arise in different contexts such as: the existence of a solid
core or by the presence of type 3A superfluid \cite{kippen}, pion condensation \cite{sawyer} or different kinds of phase transitions \cite{sokolov}. The presence of anisotropy introduces several features in the matter distribution, e.g. if we have a positive anisotropy factor $\Delta=p_{t}-p_{r}>0$, the stellar system experiences an a repulsive force (attractive in the case of negative anisotropy factor) that counteracts the gravitational gradient. Hence it allows the construction of more compact objects
when using anisotropic fluid than when using isotropic fluid \cite{mehra,harko1,Ivanov}. Furthermore, a positive anisotropy factor enhances the stability and equilibrium of the system. On the other hand, matter tensor containing
anisotropy, must be consistent with physical requirements for astrophysical applications \cite{maurya4}.\\
Under the above background our motivation in the present paper is to extent isotropic solutions to the Einstein field equations to an anisotropic scenario. Specifically we have extended Durgapal's fifth model \cite{durgapal} to the anisotropic domain.
To do so, we have applied a novel and systematic approach which opens up new possibilities for studies of compact stellar configurations that include anisotropic matter. This method was originally proposed in the context of the Randall-Sundrum braneworld \cite{randall,randall1} and
was designed to deform the standard Schwarzschild solution \cite{Ovalle4,Ovalle5}. Basically this scheme works decoupling gravitational sources through the so called minimal geometric deformation (MGD hereinafter) \cite{Ovalle,Ovalle9}. An extensive treatment of this remarkable method is given at references \cite{Ovalle1,Ovalle2,Ovalle3,Ovalle6,Ovalle7,Ovalle8} and some recent applications can be found in several frameworks e.g. purely anisotropic matter distribution \cite{Gabbanelli,camilo,Tello}, anisotropic Einstein-Maxwell system \cite{sharif,Tello1} and black holes \cite{Ovalle2018,contreras}.\\
So, this work is organized as follows: Section \ref{section2} is devoted to the construction of the isotropic extension of the Durgapal's fifth model   to an anisotropic scenario, in section \ref{section3} we analyze all the necessary requirements that an anisotropic solution of the Einstein field equations must meet to be physically admissible. Finally in section \ref{section4} some conclusions are reported.

\section{Anisotropic Durgapal's fifth model}\label{section2}

\begin{figure}[H]
\centering
\includegraphics[scale=1.3]{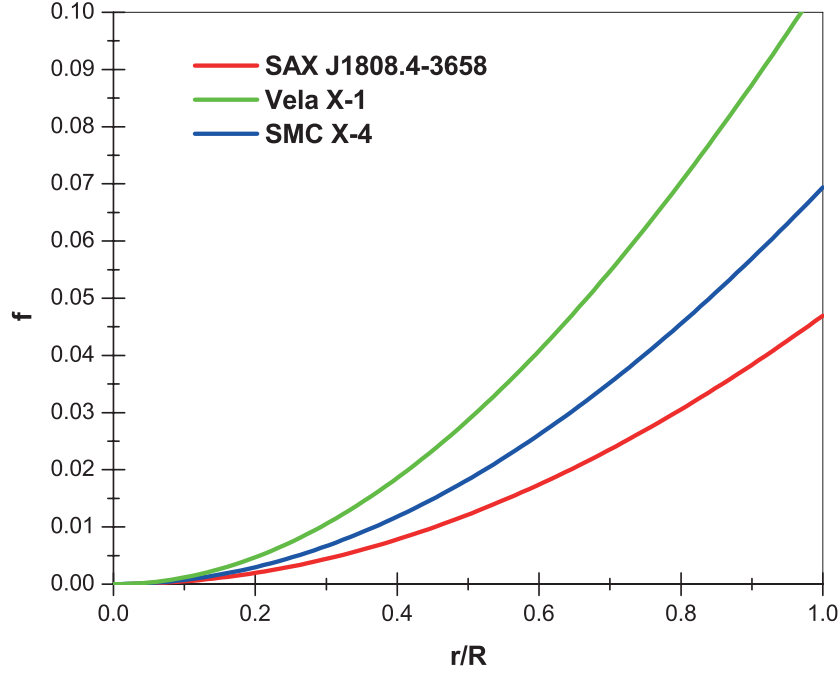}
\caption{The increasing behaviour of the deformation function $f(r)$ against the fractional radial coordinate.}
\label{function}
\end{figure}

Durgapal's fifth solution \cite{durgapal} to Einstein's field equations in Schwarzschild like coordinates is described by the following metric 
\begin{equation}\label{Durg}
\begin{split}
ds^{2}=A\left(1+Cr^{2}\right)^{5}dt^{2}-\Bigg(\frac{1-Cr^{2}\left(309-54Cr^{2}+8C^{2}r^{4}\right)}{112\left(1+Cr^{2}\right)^{3}}+\frac{BCr^{2}}{\left(1+6Cr^{2}\right)^{1/3}\left(1+Cr^{2}\right)^{3}}\Bigg)^{-1}&\\-r^{2}\left( d\theta^{2}+\sin ^{2}\theta d\phi ^{2}\right).  
\end{split}
\end{equation}
It describes a spherically symmetric and static configuration associated to an isotropic matter distribution \i.e. equal radial and transverse pressures $p_{r}=p_{t}$. Specifically the thermodynamical quantities (density and pressure) that characterize this model are
\begin{equation}\label{isodensity}
\begin{split}
\tilde{\rho}(r)=\frac{C}{8\pi}\Bigg[\frac{720C^{4}r^{8}+2820C^{3}r^{6}+540^{2}r^{4}+11625^{2}+1935}{112\left(1+Cr^{2}\right)^{4}} +\frac{\left(22C^{2}r^{4}-11Cr^{2}-3\right)B}{\left(1+6Cr^{2}\right)^{1/3}\left(1+Cr^{2}\right)^{4}}\Bigg]
\end{split}    
\end{equation}

\begin{equation}\label{isopressure}
\begin{split}
\tilde{p}(r)=-\frac{C}{8\pi}\Bigg[\frac{200C^{3}r^{6}+1050C^{2}r^{4}+4125Cr^{2}-475}{112\left(1+Cr^{2}\right)^{4}} -\frac{\left(11Cr^{2}+1\right)B}{\left(1+6Cr^{2}\right)^{1/3}\left(1+Cr^{2}\right)^{4}}\Bigg].  
\end{split}    
\end{equation}
As was pointed out earlier, one of the essential features of anisotropic models is the inequality between radial and tangential pressures \i.e. $p_{r}\neq p_{t}$. To reach this condition in the present model, the starting point is to introduce an extra gravitational source which in principle can be e.g a scalar, vectorial or tensorial field. This extra source is coupled to the energy-momentum tensor associated to the seed solution through a dimensionless coupling constant $\alpha$. Explicitly it reads 
\begin{equation}\label{effectivestresstensor}
{T}_{\mu\nu}=\tilde{T}_{\mu\nu} + \alpha \theta_{\mu\nu},   
\end{equation}
where $\tilde{T}_{\mu\nu}$ corresponds to a perfect fluid given by (\ref{isodensity})-(\ref{isopressure}). So, putting together the expressions (\ref{Durg}) and (\ref{effectivestresstensor}) the Einstein field equations are 
\begin{eqnarray}\label{effectivedensity}
8\pi {\rho}&=&\frac{1}{r^2}-e^{-\lambda}\left(\frac{1}{r^2}-\frac{\lambda^{\prime}}{r}\right)\\\label{effectiveradialpressure}
8\pi {p}_{r}&=&-\frac{1}{r^2}+e^{-\lambda}\left(\frac{1}{r^2}-\frac{\nu^{\prime}}{r}\right)\\\label{effectivetangentialpressure}
8\pi {p}_{t}&=&\frac{1}{4}e^{-\lambda}\left(2\nu^{\prime\prime}+\nu^{\prime2}-\lambda^{\prime}\nu^{\prime}+2\frac{\nu^{\prime}-\lambda^{\prime}}{r}\right).
\end{eqnarray}
The primes denote differentiation with respect to the radial coordinate $r$. From now on relativistic
geometrized units are employed, that is $c=G=1$. Bianchi's identities
\begin{equation}\label{bianchi}
\nabla^{\mu}T_{\mu\nu}=0    
\end{equation}
invokes the following conservation law
\begin{equation}
\tilde{p}^{\prime}+\frac{\nu^{\prime}}{2}\left(\tilde{p}+\tilde{\rho}\right)-\alpha\big[\left(\theta^{r}_{r}\right)^{\prime}  
+\frac{\nu^{\prime}}{2}\left(\theta^{r}_{r}-\theta^{t}_{t}\right) 
+\frac{2}{r}\left(\theta^{r}_{r}-\theta^{\varphi}_{\varphi}\right)\big]=0,
\end{equation}
being the above expression a linear combination of the equations (\ref{effectivedensity})-(\ref{effectivetangentialpressure}). Moreover, ${\rho}$, ${p}_{r}$ and ${p}_{t}$ represent the effective density, the effective radial pressure and the effective tangential pressure respectively, that are given by
\begin{eqnarray}\label{effecrho}
{\rho}&=&\tilde{\rho}+\alpha \theta^{ t}_{t}\\\label{effecpr}
{p}_{r}&=&\tilde{p}-\alpha \theta^{r}_{r}\\ \label{effecpt}
{p}_{t}&=& \tilde{p}-\alpha \theta^{ \varphi}_{\varphi}.
\end{eqnarray}
The presence of the $\theta$-term clearly introduces an anisotropy if $\theta^{r}_{r}\neq \theta^{\varphi}_{\varphi}$. Thus the effective anisotropy is defined as  
\begin{equation}\label{anisotropy}
\Delta\equiv {p}_{t}-{p}_{r}=\alpha\left(\theta^{r}_{r}- \theta^{\varphi}_{\varphi}\right).  
\end{equation}
Solve the system of equations (\ref{effectivedensity})-(\ref{effectivetangentialpressure}) is not an easy task. In order to tackle it we will employ the gravitational decoupling via the MGD approach \cite{Ovalle}. This method consists in deforming the metric potentials $e^{\nu(r)}$ and  $e^{\lambda(r)}$ through a linear mapping given by
\begin{eqnarray}\label{deformationnu}
e^{\nu(r)}&\mapsto& e^{\nu(r)}+\alpha h(r) \\ \label{deformationlambda}
e^{-\lambda(r)}&\mapsto& \mu(r)+\alpha f(r),
\end{eqnarray}
where $h(r)$ and $f(r)$ are the corresponding deformations. It's worth mentioning that the foregoing deformations are purely radial functions, this feature ensures the spherical symmetry of the solution. The so called MGD corresponds to set $h(r)=0$ or $f(r)=0$, in this case the deformation will be done only on the radial component, remaining the temporal one unchanged (it corresponds to set $h(r)=0$). Then the anisotropic sector $\theta_{\mu\nu}$ is totally contained in the radial deformation (\ref{deformationlambda}). After replacing (\ref{deformationlambda}) into the system of equations  (\ref{effectivedensity})-(\ref{effectivetangentialpressure}), it is decoupled in two systems of equations. The first one corresponds to $\alpha=0$ it means perfect fluid matter distribution 
\begin{eqnarray}\label{ro1}
8\pi\tilde{\rho}&=&\frac{1}{r^{2}}-\frac{\mu}{r^{2}}-\frac{\mu^{\prime}}{r}\\\label{p1}
8\pi \tilde{p}&=&-\frac{1}{r^{2}}+\mu\left(\frac{1}{r^{2}}+\frac{\nu^{\prime}}{r}\right)\\\label{p2}
8\pi \tilde{p}&=&\frac{\mu}{4}\left(2\nu^{\prime\prime}+\nu^{\prime2}+2\frac{\nu^{\prime}}{r}\right)+\frac{\mu^{\prime}}{4}\left(\nu^{\prime}+\frac{2}{r}\right),
\end{eqnarray}
along with the conservation equation
\begin{equation}\label{conservde}
\tilde{p}^{\prime}+\frac{\nu^{\prime}}{2}\left(\tilde{\rho} +\tilde{p}\right)=0,    
\end{equation}
it is a linear combination of the equations (\ref{ro1})-(\ref{p2}). The other set of equations corresponds to the $\theta$ sector
\begin{eqnarray}\label{cero}
8\pi\theta^{t}_{t}&=&-\frac{f}{r^{2}}-\frac{f^{\prime}}{r} 
\\ \label{one}  
8\pi\theta^{r}_{r}&=&-f\left(\frac{1}{r^{2}}+\frac{\nu^{\prime}}{r}\right)  \\  \label{dos}
8\pi\theta^{\varphi}_{\varphi}&=&-\frac{f}{4}\left(2\nu^{\prime\prime}+\nu^{\prime2}+2\frac{\nu^{\prime}}{r}\right)-\frac{f^{\prime}}{4}\left(\nu^{\prime}+\frac{2}{r}\right). \label{tres}
\end{eqnarray}
The corresponding conservation equation $\nabla^{\nu}\theta_{\mu\nu}=0$ then yields to
\begin{equation}\label{conservationtheta}
\left(\theta^{r}_{r}\right)^{\prime}-\frac{\nu^{\prime}}{2}\left(\theta^{t}_{t}-\theta^{r}_{r}\right)-\frac{2}{r}\left(\theta^{\varphi}_{\varphi}-\theta^{r}_{r}\right)=0, 
\end{equation}
this expression is a linear combination of the quasi-Einste-
in equations. At this point it is remarkable to note that both the isotropic and the anisotropic sectors are individually conserved, it means that both systems interact only gravitationally.\\ 
From the equation (\ref{Durg}) we have
\begin{equation}\label{nu}
e^{\nu(r)}=A\left(1+Cr^{2}\right)^{5}
\end{equation}
\begin{equation}\label{mu}
\mu(r)=\frac{1}{\left(1+Cr^{2}\right)^{3}}\Bigg[ \frac{1-Cr^{2}\left(309-54Cr^{2}+8C^{2}r^{4}\right)}{112}  +\frac{BCr^{2}}{\left(1+6Cr^{2}\right)^{1/3}}\Bigg],
\end{equation}
these expressions together with (\ref{isodensity}) and (\ref{isopressure}) solve the system of equations (\ref{ro1})-(\ref{p2}). On the other hand to solve the set of equations (\ref{cero})-(\ref{dos}) one needs extra information. This additional information can be for example some constraints on $\theta_{\mu\nu}$ or a suitable expression for $f(r)$. In this case we have chosen an expression for $f(r)$ given by
\begin{equation}\label{f}
f(r)=\frac{Cr^{2}}{1+Cr^{2}}.    
\end{equation}
The previous choice of the function $f(r)$ is in accordance with the behavior of the metric potentials, that is: positive, regular and increasing monotone functions with increasing radius within the compact object. Figure \ref{function} shows the increasing behaviour of the deformation function $f(r)$ with increasing radius. Then from equations (\ref{cero}), (\ref{one}) and (\ref{dos}) we obtain the following components for the source $\theta_{\mu\nu}$
\begin{eqnarray}\label{theta0}
\theta^{t}_{t}&=&-\frac{C}{8\pi}\frac{\left(3+Cr^{2}\right)}{\left(1+Cr^{2}\right)^{2}} \\ \label{theta1}
\theta^{r}_{r}&=&-\frac{C}{8\pi}\frac{\left(1+11Cr^{2}\right)}{\left(1+Cr^{2}\right)^{2}} \\ \label{theta2}
\theta^{\varphi}_{\varphi}&=&-\frac{C}{8\pi}\frac{\left(1+16Cr^{2}+25C^{2}r^{4}\right)}{\left(1+Cr^{2}\right)^{3}}.
\end{eqnarray}

Hence, the anisotropic Durgapal's fifth model is described by the following metric potentials
\begin{equation}\label{nu1}
e^{\nu(r)}=A\left(1+Cr^{2}\right)^{5},
\end{equation}
\begin{equation}\label{lambdamodificado}
e^{-\lambda(r)}=\frac{1}{\left(1+Cr^{2}\right)^{3}}\Bigg[ \frac{1-Cr^{2}\left(309-54Cr^{2}+8C^{2}r^{4}\right)}{112}  +\frac{BCr^{2}}{\left(1+6Cr^{2}\right)^{1/3}}\Bigg] + \frac{\alpha Cr^{2}}{1+Cr^{2}}, 
\end{equation}
and thermodinamically characterized by 
\begin{eqnarray}\label{anidensity}
\rho(r;\alpha)&=&\tilde{\rho}(r)-\frac{\alpha C}{8\pi}\frac{\left(3+Cr^{2}\right)}{\left(1+Cr^{2}\right)^{2}} \\ \label{pr}
p_{r}(r;\alpha)&=&\tilde{p}(r)+\frac{\alpha C}{8\pi}\frac{\left(1+11Cr^{2}\right)}{\left(1+Cr^{2}\right)^{2}} \\ \label{pt}
p_{t}(r;\alpha)&=&\tilde{p}(r)+\frac{\alpha C}{8\pi}\frac{\left(1+16Cr^{2}+25C^{2}r^{4}\right)}{\left(1+Cr^{2}\right)^{3}},
\end{eqnarray}
where $\tilde{\rho}(r)$ and $\tilde{p}(r)$ are given by the expressions (\ref{isodensity}) and (\ref{isopressure}). From the latter equations, the anisotropy factor is directly computed, yielding to

\begin{equation}\label{anifactorcomputed}
\Delta\left(\alpha;r\right)=\frac{\alpha}{4\pi}\frac{\left(2+7Cr^{2}\right)Cr^{2}}{\left(1+Cr^{2}\right)^{3}}.    
\end{equation}
Fig. \ref{anisotropy} shows the behaviour of the anisotropy factor $\Delta$. It
vanishes at $r=0$, that is so because at the center the effective radial and transverse pressures coincide. On the other hand, as the radius increases the values of these quantities drift apart, and therefore the anisotropy increases toward the
surface of the object.
\begin{figure}[H]
\centering
\includegraphics*[scale=1.3]{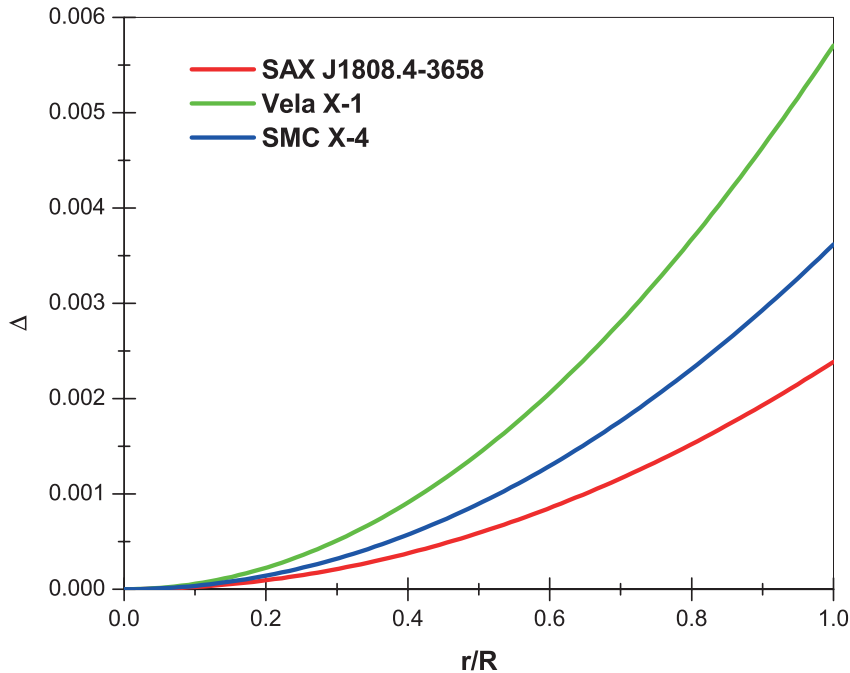}
\caption{The anisotropy factor $\Delta$ against the fractional radius for $\alpha=0.3$.}
\label{anisotropy}
\end{figure}
\subsection{Junction conditions}\label{sub1}
In order to find the arbitrary constants $A$, $B$ and $C$ we must match our interior solution (\ref{nu1})-(\ref{lambdamodificado}) to
the exterior Schwarzschild solution at the boundary of the star. The line element of the exterior Schwarzschild solution \cite{schwar} is given by
\begin{equation}\label{Schwarzschild}
ds^{2}=\left(1-\frac{2M}{r}\right)dt^{2}-\left(1-\frac{2M}{r}\right)^{-1}dr^{2} 
-r^{2}d\Omega^{2},  
\end{equation}
where $d\Omega^{2}\equiv\sin^{2}\theta d\phi^{2}+d\theta^{2}$. For this purpose we will use the Israel-Darmois junction conditions \cite{Israel,darmois}. These conditions require the continuity of the metric potentials $e^{\nu(r)}$ and $e^{\lambda(r)}$ across the surface $\Sigma$ of the compact object defined by $r=R$ (It is known as the first fundamental form). Then we have
\begin{equation}\label{match1}
A\left(1+CR^{2}\right)^{5}=1-\frac{2\tilde{M}}{R}
\end{equation}
\begin{equation}\label{match2}
\begin{split}
\frac{1}{\left(1+CR^{2}\right)^{3}}\Bigg[ \frac{1-CR^{2}\left(309-54CR^{2}+8C^{2}R^{4}\right)}{112} +\frac{BCR^{2}}{\left(1+6CR^{2}\right)^{1/3}}\Bigg] + \frac{\alpha CR^{2}}{1+CR^{2}}=1-\frac{2\tilde{M}}{R}.
\end{split}
\end{equation}
The other condition is the second fundamental form $\left[T_{\mu\nu}r^{\nu}\right]_{\Sigma}=0$, where $r^{\nu}$ is a unit vector projected in the radial direction. So the second fundamental form leads to 
\begin{equation}\label{secondfundamental}
p_{r}(R)=\left(\tilde{p}-\alpha\theta^{r}_{r}\right)(R)=0,    
\end{equation}
regarding that the Schwarzschild exterior solution describes a vacuum space-time. The equation (\ref{secondfundamental}) provides us the following expression for the constant $B$
\begin{equation}\label{B}
\begin{split}
B=-\frac{(1+6CR^2)^{1/3}}{{112(1+11CR^2)}}\bigg(1232R^6\alpha C^3-200R^6C^3 +2576R^4\alpha C^2 -1050R^4 C^2+1456R^2\alpha C & \\-4125R^2C+112\alpha+475\bigg).
\end{split}
\end{equation}
The equations (\ref{match1}), (\ref{match2}) and (\ref{B})  are the necessary and sufficient conditions to determine the constants $A$, $B$ and $C$. In addition, the values of the mass $\tilde{M}$ and the radius $R$ have been established based on the obtained data from some strange star candidates \cite{tapa}.In table \ref{table1} are displaying the values of the constant parameters $A$, $B$ and $C$ for each strange star candidate.
\section{Physical analysis}\label{section3}
In this section we will analyze the basic requirements that all anisotropic solution to Einstein field equations must fulfill in order to be an admissible physical model describing stellar interiors \cite{Herrera}.
\subsection{Regularity}
\begin{figure}[H]
\centering
\includegraphics[scale=1.3]{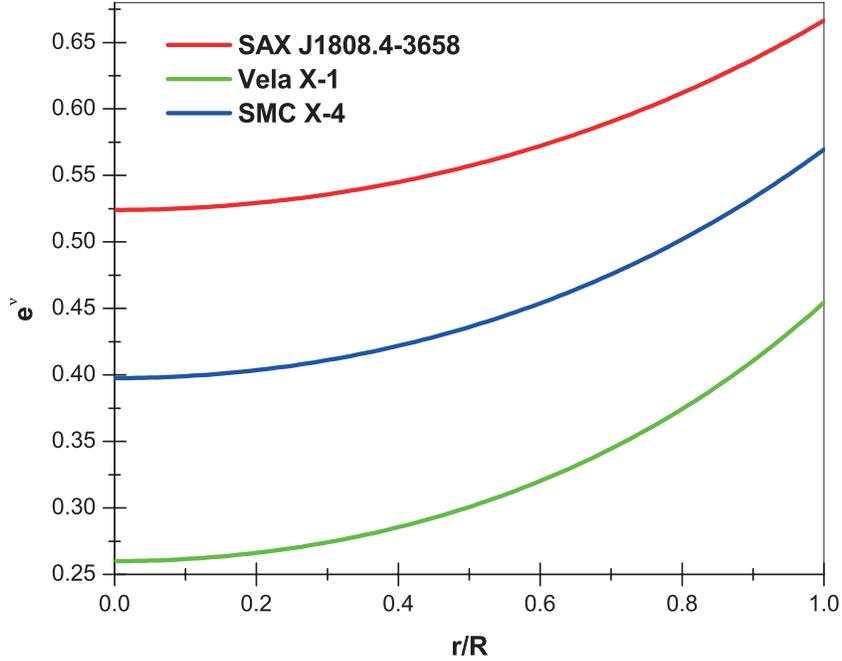}
\caption{The temporal metric component versus the dimensionless radius for $\alpha=0.3$.}
\label{gtt}
\end{figure}

\begin{figure}
\centering
\includegraphics[scale=1.3]{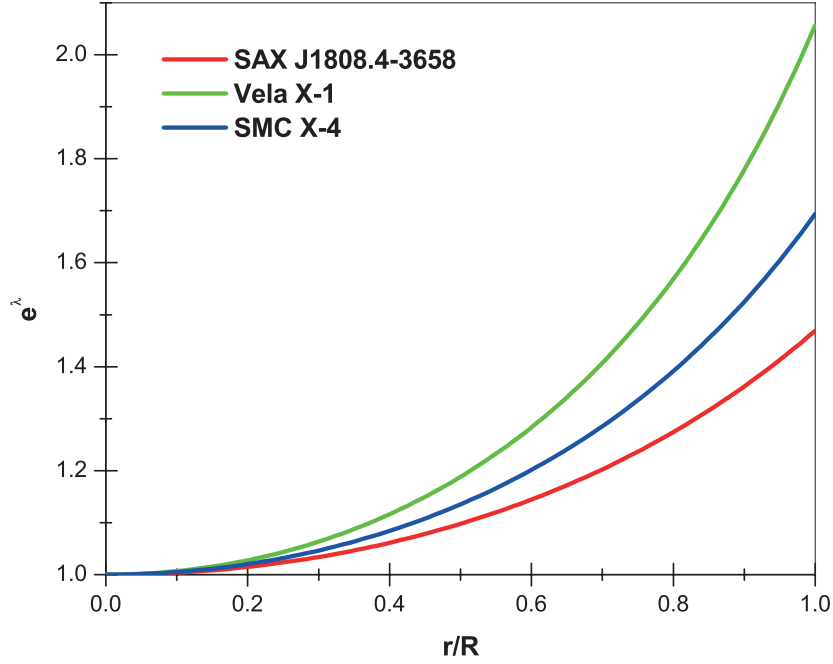}
\caption{The deformed radial metric component versus the dimensionless radius for $\alpha=0.3$.}
\label{grr}
\end{figure}

The departure point is to analyze if the model is free from  physical and geometric singularities and non zero positive values of $e^{\nu(r)}$ and $e^{\lambda(r)}$ i.e $(e^{\lambda(r)})|_{r=0}=1$ and $(e^{\nu(r)})|_{r=0}>0$. It is clear from expressions (\ref{nu1}) and (\ref{lambdamodificado}) that $e^{\lambda(0)}=1$ and $e^{\nu(0)}=A$, obviously it demands $A>0$. Figures \ref{gtt} and \ref{grr} show the behaviour of both metric functions against the dimensionless radial coordinate.

\begin{figure}[H]
\centering
\includegraphics[scale=1.3]{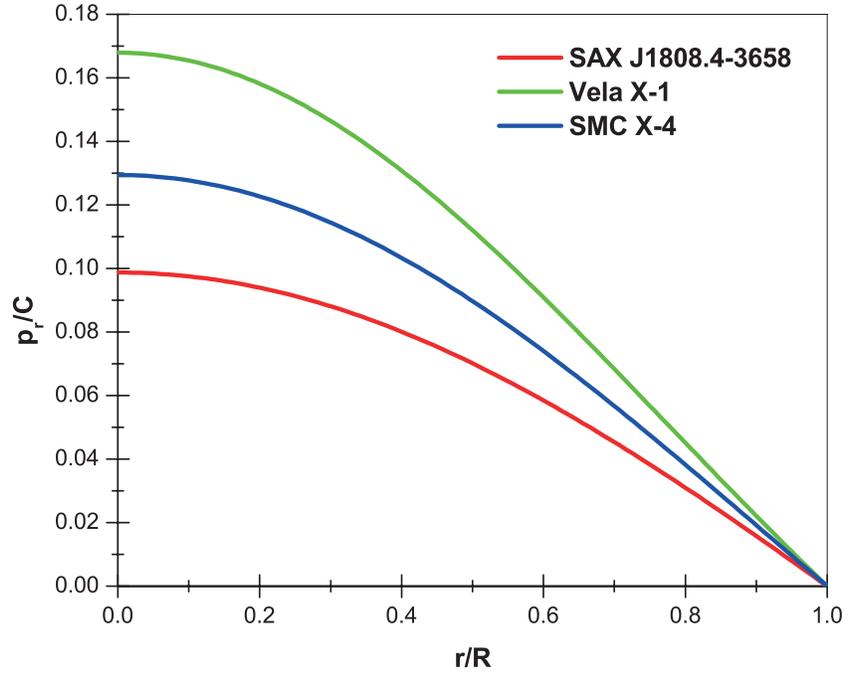}
\caption{The dimensionless radial pressure against the fractional radius for $\alpha=0.3$.}
\label{radialpressures}
\end{figure}

\begin{figure}[H]
\centering
\includegraphics[scale=1.3]{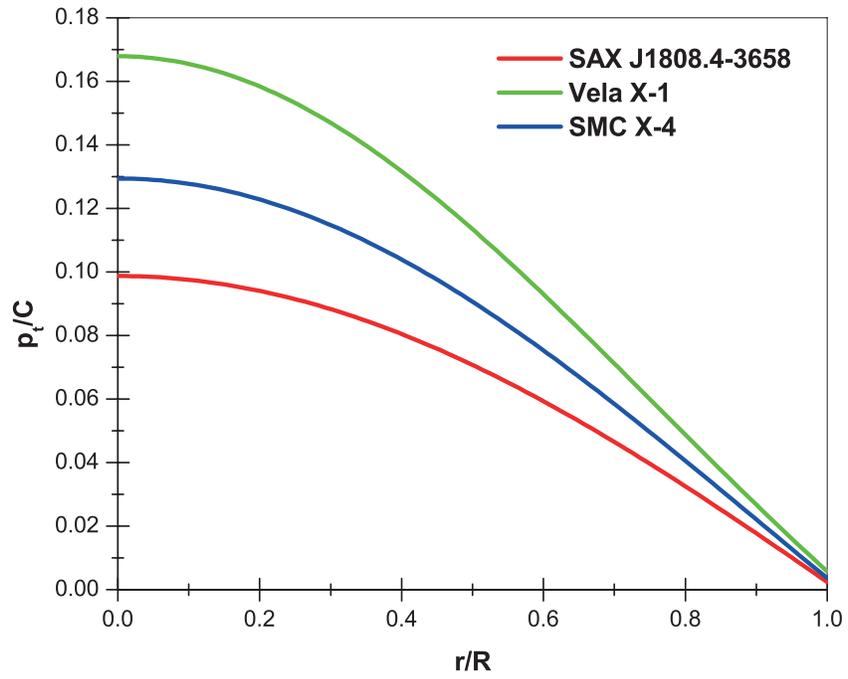}
\caption{The dimensionless tangential pressure against the fractional radius for $\alpha=0.3$.}
\label{tangentialpressures}
\end{figure}

\begin{figure}[H]
\centering
\includegraphics[scale=1.3]{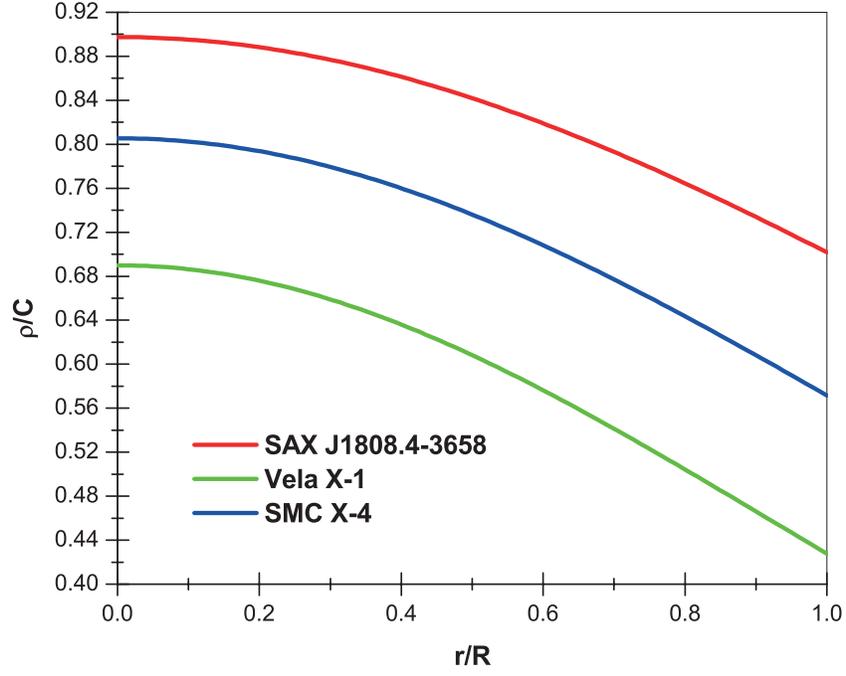}
\caption{The dimensionless density against the fractional radius for $\alpha=0.3$.}
\label{densities}
\end{figure}

\begin{figure}[H]
\centering
\includegraphics*[scale=1.5]{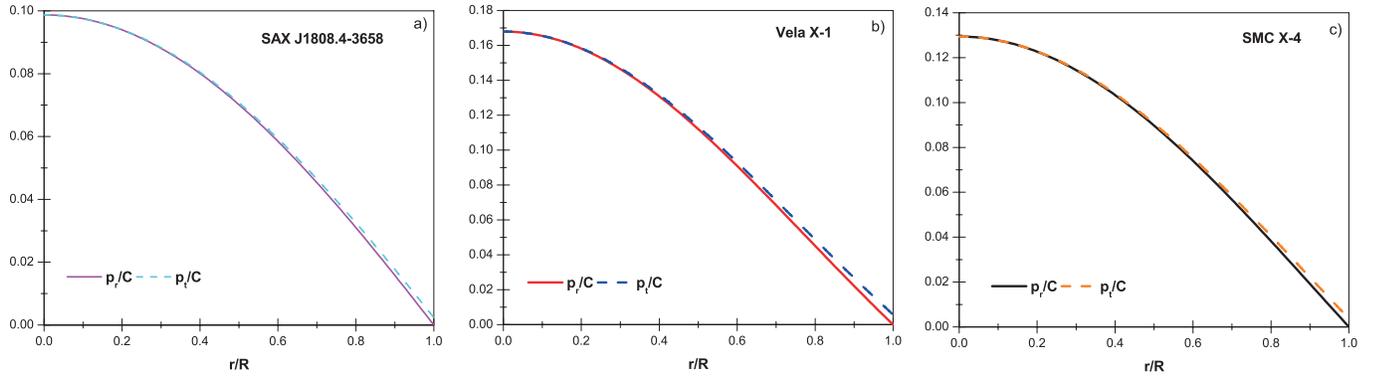}
\caption{Comparison between the radial and the tangential pressures for $\alpha=0.3$.}
\label{pressures}
\end{figure}

\begin{figure}[H]
\centering
\includegraphics[scale=1.3]{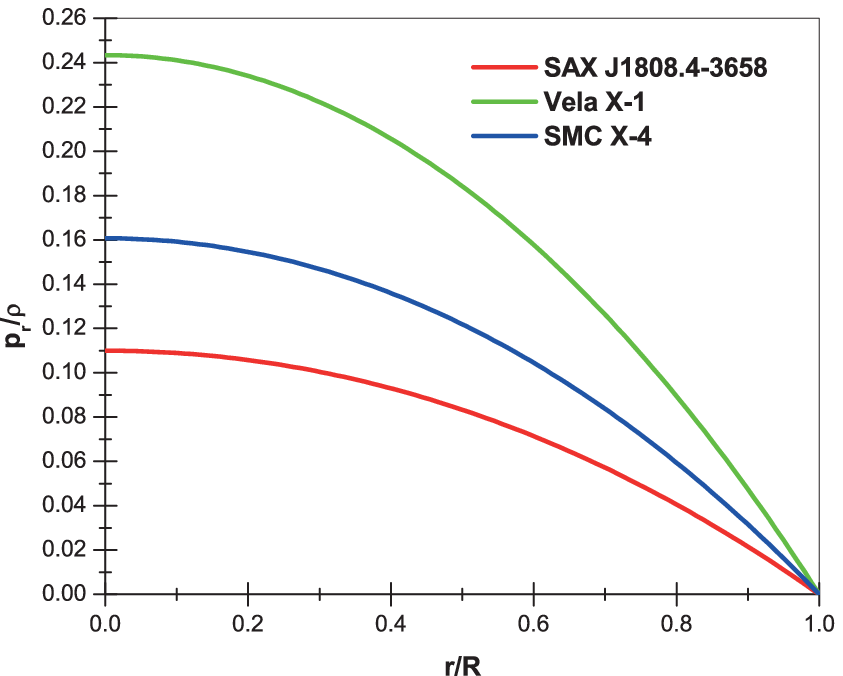}
\caption{Zeldovich's condition in the radial direction versus the fractional radius for $\alpha=0.3$.}
\label{zeldovichradial}
\end{figure}

\begin{figure}[H]
\centering
\includegraphics[scale=1.3]{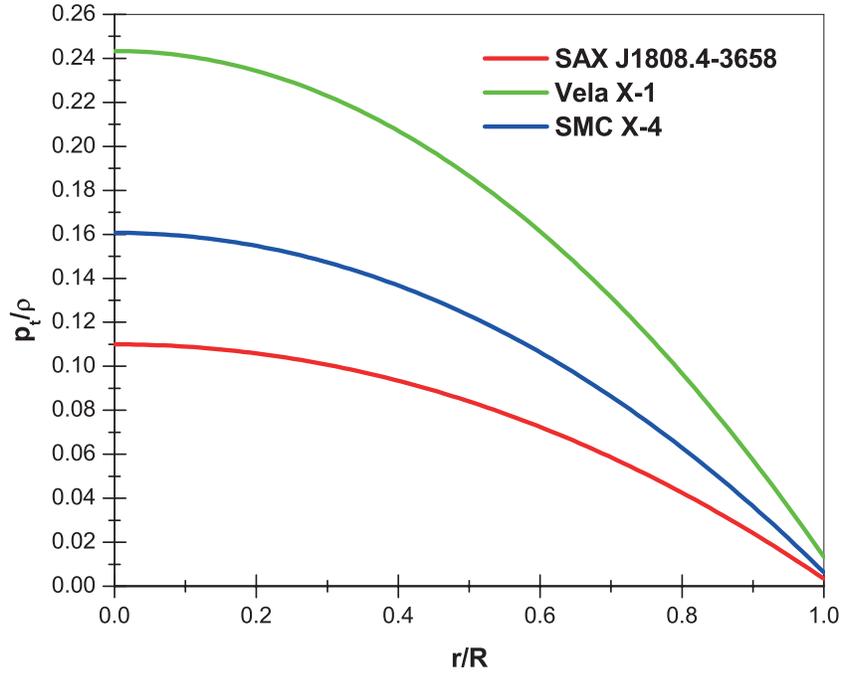}
\caption{Zeldovich's condition in the tangential direction versus the fractional radius for $\alpha=0.3$.}
\label{zeldovichtangential}
\end{figure}

Another important aspect is related to the behaviour of the effective thermodynamic observables $\rho(r)$, $p_r(r)$ and $p_{t}(r)$ within the stellar configuration. They must be positive and monotonically decreasing functions as they approach the to boundary of the compact object and their maximum values must be attain at the center $r=0$. Evaluating (\ref{pr}) and (\ref{pt}) at $r=0$ we obtain
\begin{equation}\label{pressureatcenter}
p_{r}(0)=p_{t}(0)=\frac{(475+112B)C}{112}+\alpha C >0,   
\end{equation}
then
\begin{equation}\label{upperB}
B>-\frac{475+112\alpha}{112}.   
\end{equation}
Moreover, to satisfy Zeldovich condition at the interior $p_{r}/\rho$ at center must be $\leq 1$. Therefore
\begin{equation}\label{lowerB}
B\leq \frac{1460-448\alpha}{448}.   
\end{equation}
In order to ensure the positiveness of $p_{r}$ and $p_{t}$ inside the distribution, from (\ref{pressureatcenter}) the constant $C$ must be positive. On the other hand (\ref{upperB}) and (\ref{lowerB}) imply a constraint over the constant $B$ given by 
\begin{equation} -\frac{475+112\alpha}{112}   <B\leq\frac{1460-448\alpha}{448}.
\end{equation}
For the coupling constant $\alpha$ we have
\begin{equation}\label{45}
0<\alpha<1,    
\end{equation}
throughout the study we have fixed $\alpha=0.3$. The bound given for (\ref{45}) leads to $p_{t}>p_{r}\Rightarrow \Delta> 0$ which represents a force due to the anisotropy directed outwards. Therefore we should have more massive and compact configurations \cite{mehra,harko1}. We can observe from figures \ref{radialpressures}, \ref{tangentialpressures} and \ref{densities} that $p_{r}$, $p_{t}$ and $\rho$ are monotonically decreasing functions with increasing radius and attain their maximum values at center of the star. 
Besides panels $a$, $b$ and $c$ in figure \ref{pressures} shown how the effective radial pressure $p_{r}$ and the effective tangential $p_{t}$ pressure drift apart. The Zeldovich's condition is shown in figures \ref{zeldovichradial} and \ref{zeldovichtangential}. Clearly it is satisfied for the present model.   Table \ref{table2} exhibits the corresponding values of the central and surface density, the central pressure and the central pressure-central density ratio, for the chosen strange star candidates. 

\subsection{Causality condition}
\begin{figure}[H]
\centering
\includegraphics[scale=1.3]{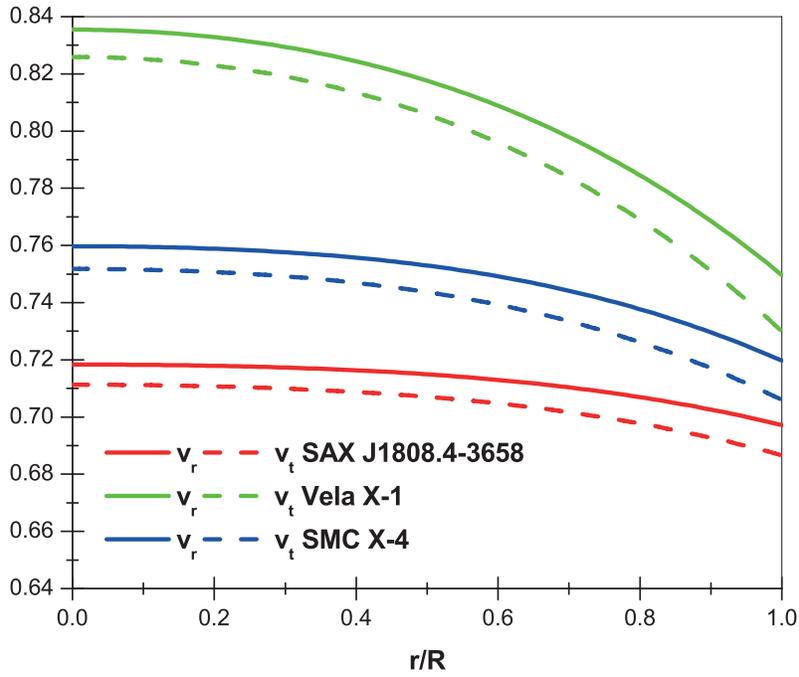}
\caption{Sound speeds against the dimensionless radius for $\alpha=0.3$.}
\label{velocities}
\end{figure}

An admissible anisotropic solution to Einstein field equations must satisfies causality condition \i.e both the radial $v_{r}$ and tangential $v_{t}$ sound speeds inside the object are less than the speed of light $c$ (in relativistic geometrized units the speed of light becomes $c=1$ ). Explicitly it reads 
\begin{eqnarray}
v_{r}(r)&=&\sqrt{\frac{dp_{r}(r)}{d\rho(r)}}\leq1 \\
v_{t}(r)&=&\sqrt{\frac{dp_{t}(r)}{d\rho(r)}}\leq1.
\end{eqnarray}
As shown in figure \ref{velocities} both speeds fulfill the above requirement. In addition, they have their maximum value at the center of the object (high density region) and decrease monotonically towards the surface (lower density region). 

\subsection{Energy conditions}
\begin{figure}[H]
\centering
\includegraphics[scale=2.0]{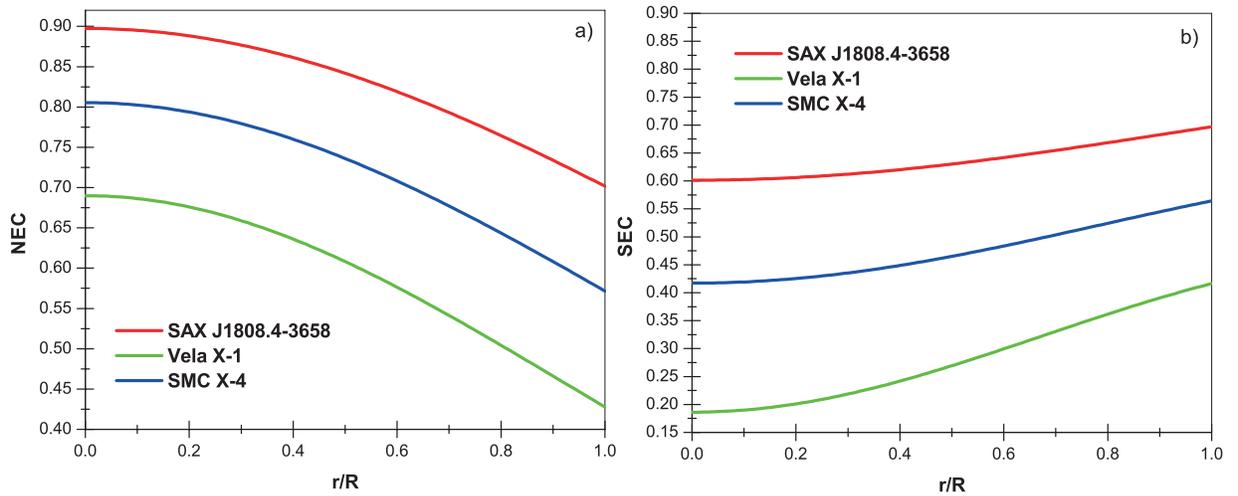}
\caption{Null energy condition panel $a)$ and Strong energy condition panel $b)$.}
\label{NECSEC}
\end{figure}

\begin{figure}[H]
\centering
\includegraphics[scale=2.0]{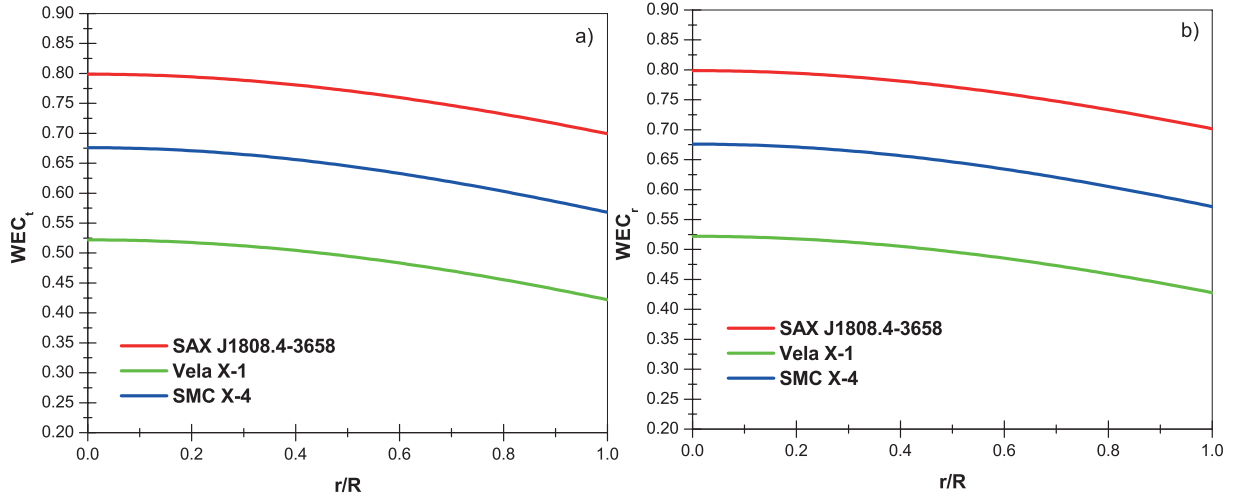}
\caption{Weak energy conditions in both tangential and radial direction, panels $a)$ and $b)$ respectively.}
\label{WEC}
\end{figure}

\begin{figure}[H]
\centering
\includegraphics[scale=2.0]{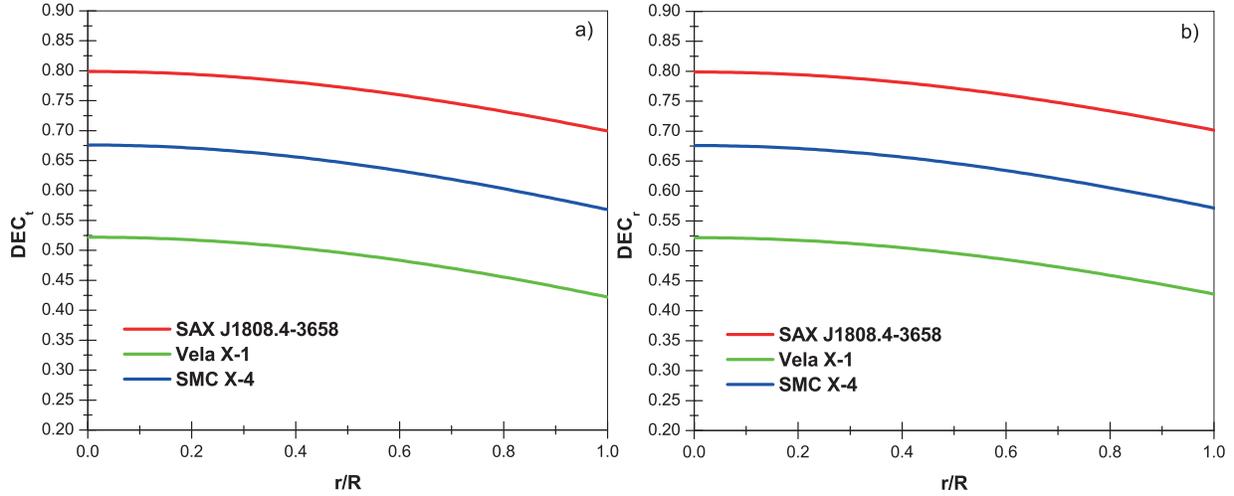}
\caption{Dominant energy conditions in both tangential and radial direction, panels $a)$ and $b)$ respectively.}
\label{DEC}
\end{figure}

Within the anisotropic matter distribution the energy should be positive. In order to ensure it, the energy-momentum tensor has to obey the null energy condition (NEC), the weak energy condition (WEC) in both tangential and radial direction, the strong energy condition (SEC) and the dominant energy conditions (DEC) in both tangential and radial direction\cite{ponceLeon1,visserbook}:
\begin{enumerate}
    \item (NEC): $\rho \geq 0$.
    \item (WEC): $\rho-p_{t}\geq 0$, $\rho-p_{r}\geq 0$ .
    \item (SEC): $\rho-2p_{t}-p_{r}\geq 0$.
    \item (DEC): $\rho -|p_{r}|\geq0$, $\rho -|p_{t}|\geq0$.
\end{enumerate}
 Figures \ref{NECSEC}, \ref{WEC} and \ref{DEC} shown that all the above inequalities are satisfied within the object. Therefore we have a well behaved energy-momentum tensor.

\subsection{Effective mass-radius ratio and redshift}
\begin{figure}[H]
\centering
\includegraphics[scale=1.3]{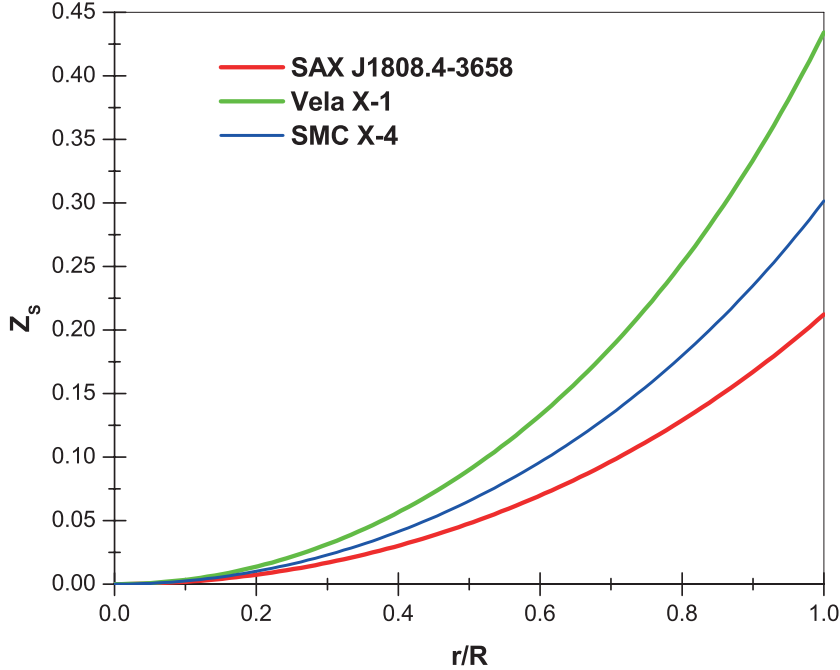}
\caption{The surface redshift against the dimensionless radius for $\alpha=0.3$.}
\label{surfacerdshift}
\end{figure}

In the study of spherically symmetric and static distributions associated with a perfect fluid, the maximum limit of the mass-radius ratio must satisfy the following upper bound $u=\tilde{M}/R<4/9$ (in the units $c=G=1$) \cite{buchdahl}. However, in the presence of an anisotropic matter distribution this limit is more general \cite{andreasson}. However, it can be obtained from the effective mass defined as \cite{harko1}
\begin{equation}
M_{eff}=4\pi\int^{R}_{0}\rho r^{2} dr=\frac{R}{2}\left[1-e^{-\lambda(R)}\right].
\end{equation}
Explicitly the effective mass $M_{eff}$ reads
\begin{equation}\label{effectivemass}
\begin{split}
M_{eff}=\frac{CR^{3}}{2\left(1+6CR^{2}\right)^{1/3}\left(1+CR^{2}\right)^{3}} \Bigg\{\Bigg[C^{2}\left(\frac{15}{14}-\alpha\right)R^{4} 
+\left(\frac{195}{56}C-2C\alpha\right)R^{2}+\frac{645}{112}-\alpha\Bigg]&\\
\times\left(1+CR^{2}\right)^{1/3}-B\Bigg\}.   
\end{split}
\end{equation}
Then the compactness factor $u$ becomes 
\begin{equation}\label{compactnessfactor}
\begin{split}
u=\frac{CR^{2}}{2\left(1+6CR^{2}\right)^{1/3}\left(1+CR^{2}\right)^{3}} \Bigg\{\Bigg[C^{2}\left(\frac{15}{14}-\alpha\right)R^{4} 
+\left(\frac{195}{56}C-2C\alpha\right)R^{2}+\frac{645}{112}-\alpha\Bigg]&\\
\times\left(1+CR^{2}\right)^{1/3}-B\Bigg\}.   
\end{split}
\end{equation}
So, the surface redshift (see Fig. \ref{surfacerdshift}) can be calculated using the compactness factor $u$ given by (\ref{compactnessfactor}), as follows
\begin{equation}
Z_{s}=\sqrt{e^{-\lambda(R)}}-1=\frac{1-\sqrt{1-2u}}{\sqrt{1-2u}}.    
\end{equation}
The presence of a positive anisotropy factor $\Delta>0$ does not impose an upper limit on the surface redshift $Z_{s}$. In distinction with what happens in the case of isotropic distributions, where the maximum value that the surface redshift $Z_{s}$ can reaches is $Z_ {s} = 4.77$ \cite{bowers}. Therefore, the surface redshift for anisotropic matter configurations is greater than its isotropic
counterpart.

\subsection{Equilibrium condition}
\begin{figure}[H]
\centering
\includegraphics[scale=1.3]{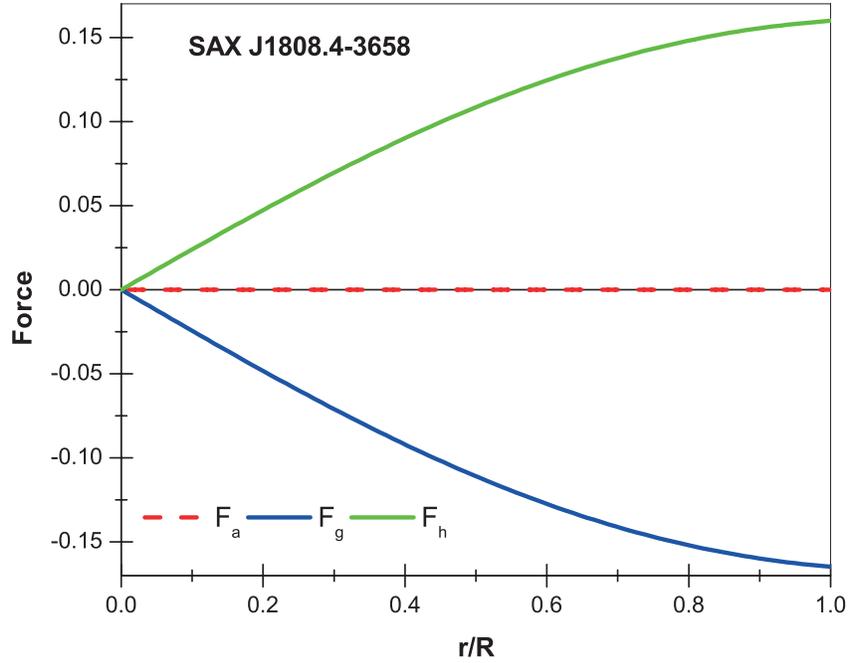}
\caption{TOV equation for the strange star candidate SAX J1808.4-3658 versus the fractional radial coordinate for $\alpha=0.3$.}
\label{SAX}
\end{figure}

\begin{figure}[H]
\centering
\includegraphics[scale=1.3]{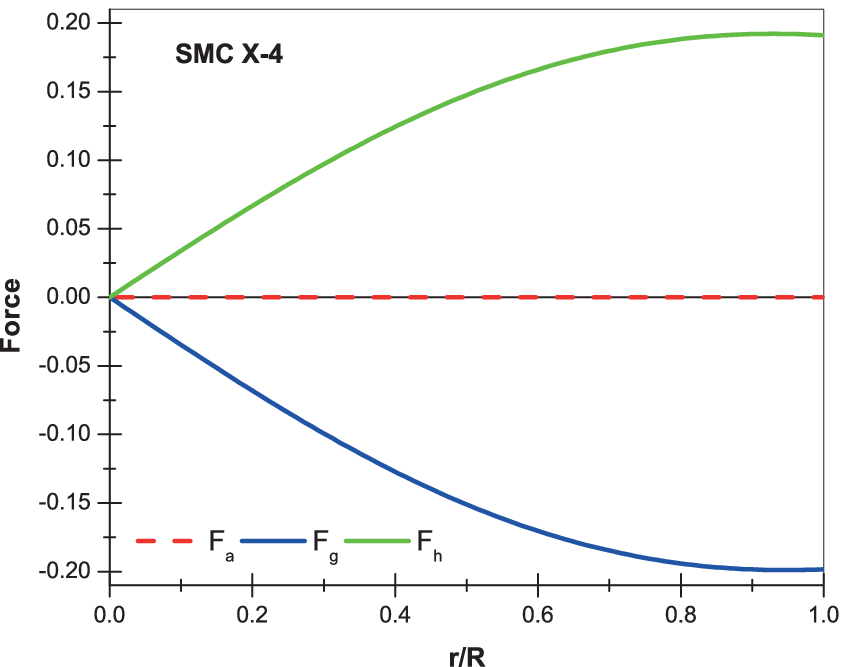}
\caption{TOV equation for the strange star candidate SMC X-4 versus the fractional radial coordinate for $\alpha=0.3$.}
\label{SMC}
\end{figure}

\begin{figure}[H]
\centering
\includegraphics[scale=1.3]{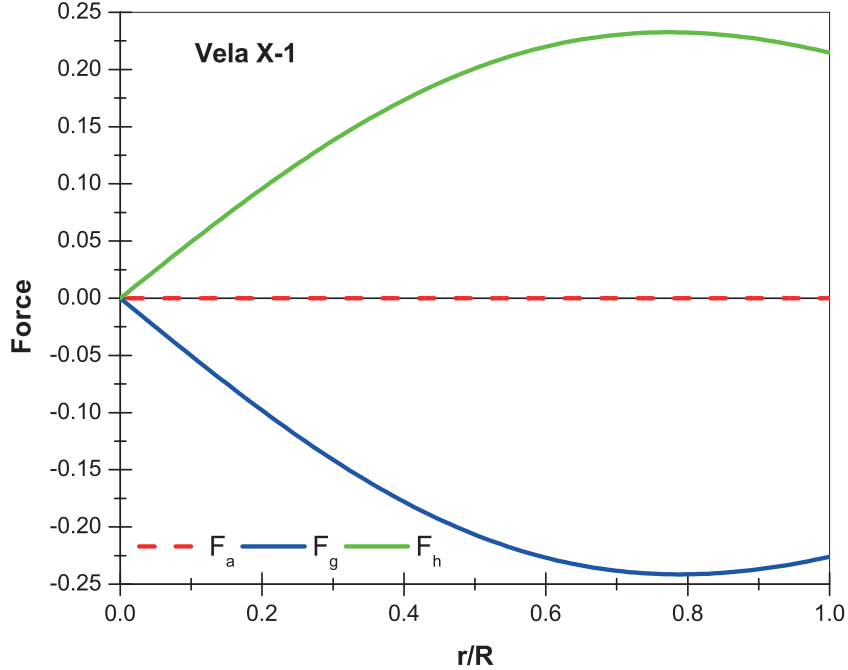}
\caption{TOV equation for the strange star candidate Vela X-1 versus the fractional radial coordinate for $\alpha=0.3$.}
\label{vela}
\end{figure}

The equilibrium of the system lies on the Tolman-Oppenheimer-Volkoff (TOV) equation \cite{tolman,oppenheimer}. In this case the equilibrium of the anisotropic fluid sphere is under three different forces. This forces are the hidrostatic force $F_{h}$, the gravitational force $F_{g}$ and the anisotropic repulsive force $F_{a}$ introduced by the presence of a positive anisotropy factor $\Delta$. In fact, the hidrostatic force $F_{h}$ and the anisotropic repulsive force $F_{a}$ counterbalance the gravitational force $F_{g}$. Therefore, the collapse of the compact object to a point singularity may be avoided during the gravitational collapse. In conclusion, the presence of anisotropies within the stellar configuration enhance the equilibrium of the system. \\
Then the TOV equation describing the equilibrium condition for an anisotropic 
fluid distribution is given by

\begin{equation}\label{TOV}
-\underbrace{\frac{\nu^{\prime}}{2}\left(\rho+p_{r}\right)}_{F_{g}} - \underbrace{\frac{dp_{r}}{dr}}_{F_{h}}+\underbrace{\frac{2}{r}\Delta}_{F_{a}}=0.
\end{equation}
We can observe from figures \ref{SAX}, \ref{SMC} and \ref{vela} that the model is in equilibrium under the mentioned forces.

\subsection{Stability conditions}
\begin{figure}[H]
\centering
\includegraphics[scale=1.3]{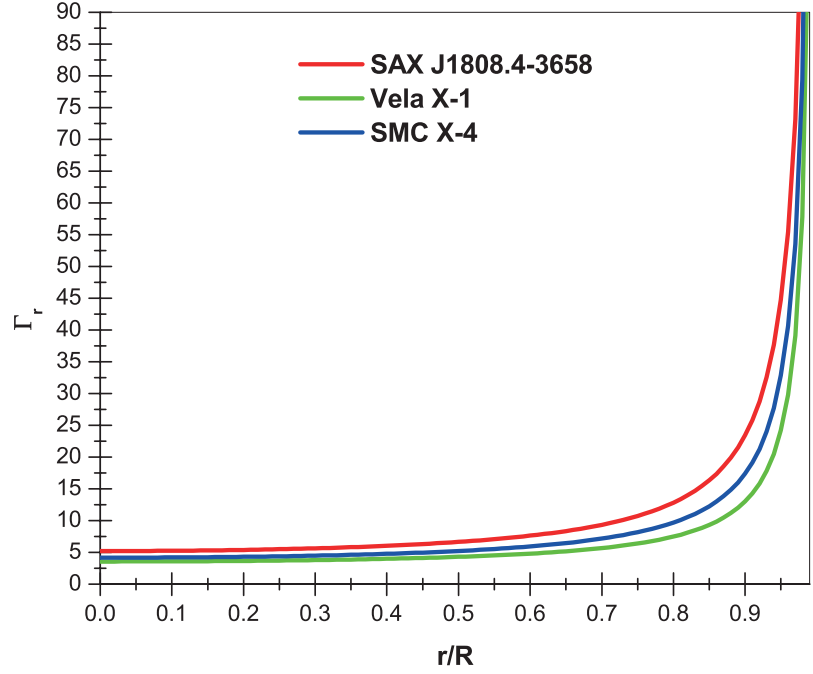}
\caption{Behaviour of adiabatic index versus the dimensionless radius for $\alpha=0.3$.}
\label{adiabaticindex}
\end{figure}

\begin{figure}[H]
\centering
\includegraphics[scale=1.2]{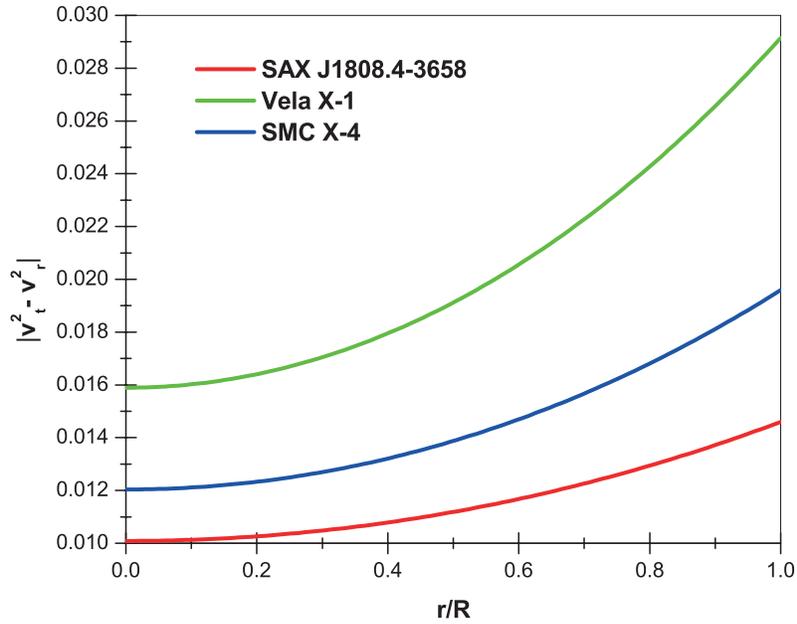}
\caption{Variation of the absolute value of square of sound velocity
with respect to fractional radius corresponding to $\alpha=0.3$.}
\label{absvelocities}
\end{figure}

\begin{figure}[H]
\centering
\includegraphics[scale=1.3]{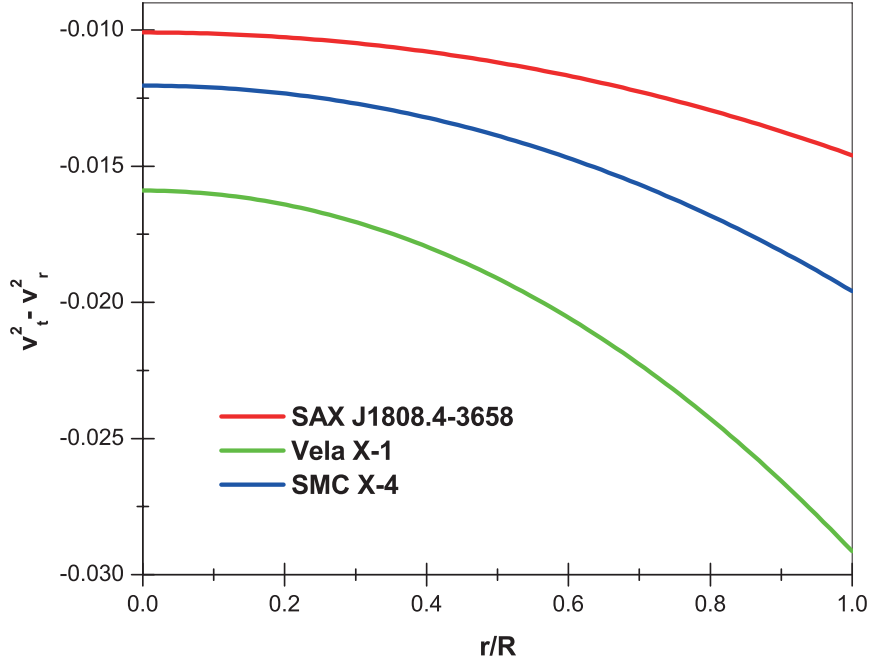}
\caption{Behaviour of $v^{2}_{t}-v^{2}_{r}$ against the fractional radius for $\alpha=0.3$.}
\label{squarevelocities}
\end{figure}

Another relevant aspect in the study of anisotropic fluid spheres which is enhanced by the presence of anisotropies in the matter distribution is the stability. In order to verify if the present model is stable we analyze both the relativistic adiabatic index $\Gamma$ \cite{bondi,heintz} and the square of sound speeds behaviour inside the compact object \cite{abreu}.\\
It is well known from the studies about Newtonian isotropic fluid spheres that the collapsing condition correspond to $\Gamma<4/3$. In distinction with the relativistic counterpart this condition becomes \cite{chan1,chan2}
\begin{equation}\label{adibatic}
\Gamma<\frac{4}{3}+\left[\frac{1}{3}\kappa\frac{\rho_{0}p_{r0}}{|p^{\prime}_{r0}|}r+\frac{4}{3}\frac{\left(p_{t0}-p_{r0}\right)}{|p^{\prime}_{r0}|r}\right]_{max},    
\end{equation}
where $\rho_{0}$, $p_{r0}$ and $p_{t0}$ are the initial density, radial and tangential pressure when the fluid is in static equilibrium. The second term in the right hand side represents the relativistic corrections to the Newtonian perfect fluid and the third term is the contribution due to anisotropy. It is clear from (\ref{adibatic}) that in the case of a non-relativistic matter distribution and taking $p_{r}$ be equal to $p_{t}$ \i.e $\Delta=0$, the bracket vanishes and we recast the collapsing Newtonian limit $\Gamma<4/3$. On the other hand, Heintzmann and Hillebrandt \cite{heintz} showed that in the presence of a positive an increasing anisotropy factor $\Delta=p_{t}-p_{r}>0$, the stability condition for a relativistic compact object is given by $\Gamma>4/3$, this is so because positive anisotropy factor
may slow down the growth of instability. Due to the fact that gravitational collapse occurs in the radial direction, it is enough to analyze the adiabatic index in such direction. So, the adiabatic index is given by \cite{chan3}  
\begin{equation}
\Gamma_{r}=\frac{\rho+p_{r}}{p_{r}}\frac{dp_{r}}{d\rho}.    
\end{equation} 
We can see from figure \ref{adiabaticindex} that the model is in complete agreement with the condition $\Gamma_{r}>4/3$. Then the model is stable.\\
Based on the Herrera's cracking concept \cite{herrera} Abreu et. al. \cite{abreu} established another alternative to study the stability of a self-gravitating anisotropic fluid sphere. This approach states that the region is potentially stable where the radial speed $v_{r}$ of sound is greater than the 
transverse $v_{t}$ speed of sound. This implies that there is no change in sign $v^{2}_{t}-v^{2}_{r}$. The later assumption is equivalent to $0\leq|v^{2}_{t}-v^{2}_{r}|<1$.\\
We note from figure \ref{velocities} that radial speed of sound is always greater than
transverse speed of sound and also from figure \ref{absvelocities} $0\leq|v^{2}_{t}-v^{2}_{r}|<1$ everywhere inside the star. On the other hand, figure \ref{squarevelocities} shows that there in no change in sign.
These features represent that the proposed physical model is stable.

\begin{table}[H]
\caption{Constant parameters calculated for radii and mass for some strange star candidates with $\alpha=0.3$}
\label{table1}
\begin{tabular*}{\textwidth}{@{\extracolsep{\fill}}lrrrrrl@{}}
\hline
Strange star & \multicolumn{1}{c}{radii $(R)/$} & \multicolumn{1}{c}{$M/$} & \multicolumn{1}{c}{$C/$} & \multicolumn{1}{c}{$A$(dimen-}&
\multicolumn{1}{c}{$B$ (dimen-}
 \\
candidates &$(km)$& $M_{\odot}$&$(\times10^{-4}km^{-2})$& sionless)& sionless) \\
\hline
SMC X$-$4 & 8.831 & 1.29 & 9.55880 & 0.39760& -1.28845 \\
\hline
Vela X$-$1 &9.560 & 1.77 & 12.9031 & 0.26000 & -0.32103 \\ \hline
SAX J1808.4$-$3658 & 7.951 & 0.90 & 7.79019 & 0.52408 & -2.06007 \\
\hline
\end{tabular*}
\end{table}

\begin{table}[H]
\caption{Some physical parameters calculated for radii and mass for some strange star candidates with $\alpha=0.3$}
\label{table2}
\begin{tabular*}{\textwidth}{@{\extracolsep{\fill}}lrrrrrl@{}}
\hline
Strange star & \multicolumn{1}{c}{$\rho(0)/$} & \multicolumn{1}{c}{$\rho(R)/$} & \multicolumn{1}{c}{$p_{r}(0)/$} & \multicolumn{1}{c}{$p_{r}(0)/$}&
 \\
candidates &$(\times 10^{15} gcm^{-3})$& $(\times 10^{15} gcm^{-3})$&$\rho(0)$&$(\times 10^{35} dyne/ cm^{2})$\\
\hline
SMC X$-$4 & 1.03881 & 0.73732 & 0.16069 & 1.50229 \\
\hline
Vela X$-$1 &1.20371 &0.74638 &0.24337 &2.63655\\
\hline
SAX J1808.4$-$3658 & 0.94342 & 0.07376 & 0.10998 & 0.93389 \\
\hline
\end{tabular*}
\end{table}

\section{Concluding remarks}\label{section4}
Gravitational decoupling through minimal geometric deformation approach offers to us a new window to generate anisotropic solutions to Einstein field equations. Despite its simplicity this powerful tool allows to us a better understanding about self-gravitating anisotropic configurations. The advantage of this scheme lies in the fact that it splits a complicated system of equation in two simple separated set of equations, one corresponding to the usual Einstein field equations associated with an isotropic matter distribution (perfect fluid) and the second one governed by an extra gravitational source $\theta_{\mu\nu}$ which encodes the anisotropic sector (this system of equation also is known as quasi-Einstein equations, as a consequence of a missed  $-1/r^{2}$ term which avoid the matching with standard Einstein equations). In this opportunity Durgapal's fifth model solves the perfect fluid sector, and the quasi-Einstein equations are solved once an additional information is given on the components of the source $\theta_{\mu\nu}$ (this additional information may be for example some constraints type equation of state) or choose an adequate expression on the minimal geometric deformation function $f(r)$ (like in our case). It is worth mentioning that both sectors the isotropic and the anisotropic one are independently conserved, this means that there is no exchange of energy and therefore the interaction between both sectors are purely gravitational.\\
In this work only radial deformation was reported. However, temporal deformation may brings interesting results too. On the other hand, the obtained extension of Durgapal's fifth model to anisotropic domain representing some strange stars  candidates is an admissible anisotropic solution to Einstein field equations. This is so because, the solution fulfills all the requirements from the physical point of view leading to a well behaved anisotropic matter distribution.

\section{Acknowledgements}
F. Tello-Ortiz thanks the Financial support by the project ANT1756 at the Universidad de Antofagasta, Chile.

\end{document}